\documentclass[aps,prl,12pt,twocolumn,tightenlines,superscriptaddress,
amsfonts,amssymb,amsmath,byrevtex,showpacs,nofootinbib]{revtex4}
\usepackage{graphics}
\usepackage{epsfig,subfigure}

\newcounter{mnotecount}[section]
\renewcommand{\themnotecount}{\thesection.\arabic{mnotecount}}
\newcommand{\mnote}[1]
{\protect{\stepcounter{mnotecount}}$^{\mbox{\footnotesize  $
      \bullet$\themnotecount}}$ \marginpar{\raggedright\tiny
    $\!\!\!\!\!\!\,\bullet$\themnotecount: #1} }

\begin{document}
\newcommand{\dR}{\mathbb R}
\newcommand{\dC}{\mathbb C}
\newcommand{\dZ}{\mathbb Z}
\newcommand{\id}{\mathbb I}
\newcommand{\Ar}{\mathop {\rm Ar}\nolimits}
\newcommand{\sgn}{\mathop {\rm sgn}\nolimits}

\title{Energy Scale of the Big Bounce}

\author{Przemys{\l}aw Ma{\l}kiewicz}
\email{pmalk@fuw.edu.pl}
\author{W{\l}odzimierz Piechocki}
\email{piech@fuw.edu.pl} \affiliation{  Theoretical Physics
Department Institute for Nuclear Studies,
\\ Ho\.{z}a 69, 00-681 Warszawa, Poland}

\date{\today}

\begin{abstract}
We examine the nature of the cosmological big bounce (BB)
transition within the loop geometry underlying Loop Quantum
Cosmology (LQC)  at classical and quantum levels. Our canonical
quantization method is an alternative to the standard LQC. An
evolution parameter we use has a clear interpretation.  Our method
opens door for analyzes  of spectra of physical observables like
the energy density and the volume operator. We find that one
cannot determine the energy scale specific to BB by making use of
the loop geometry without an extra input from observational
cosmology.
\end{abstract}
 \maketitle

\noindent{\it 1. Introduction.} It is  commonly believed that the
unification of gravity and quantum mechanics would solve the
intriguing problem of   singularities in general relativity (GR).
Canonical quantization is the preferred method of quantization of
GR.  In the canonical formulation of GR the Hamiltonian is a sum
of first-class constraints.  There are  two ways of quantization
of such  systems \cite{PAM,NMJ} based on prescriptions: (i) `first
quantize kinematics, then impose constraints', or (ii) `first
solve classical constraints, then impose quantum rules'. The
latter case is applied in the present paper. The former one is
called the Dirac quantization and it is the quantization method
used in the so called Loop Quantum Gravity (LQG) (see, e.g.
\cite{CR,TT}) and Loop Quantum Cosmology (LQC) (see, e.g.
\cite{Ashtekar:2003hd,Bojowald:2006da}).

LQC deals with the simplest cosmological models of the universe.
It offers the resolution of the initial big-bang singularity in
the sense that the singularity is replaced  by the regular
big-bounce (BB) transition (see, e.g.
\cite{Ashtekar:2006rx,Ashtekar:2006uz,Ashtekar:2006wn,Ashtekar:2007em}).
Revealing the nature of the initial singularity is a prerequisite
for understanding  the origin of matter, non-gravitational fields
and spacetime. Thus, much importance has been  ascribed to this
result.

We define the volume operator and present preliminary results
concerning its spectrum in the physical Hilbert space.

The aim of our paper is to show that the energy scale specific to
the BB transition cannot be determined uniquely because the energy
density of matter specific to BB depends on a free parameter
labelling  the loop geometry. The determination of this scale is
of primary importance as it would help to identify  the
unification scale of gravity with quantum physics.

\noindent{\it 2. Classical dynamics.} For simplicity of exposition
we restrict ourselves to  the quantization problem of  flat
Friedmann-Robertson-Walker (FRW) model with massless scalar field.
The metric in this model reads
 \begin{equation}\label{metric}
    ds^2=-N^2(t)\,dt^2+a^2(t)\,(dx^2+dy^2+dz^2),
\end{equation}
where $a$ is the scale factor and $N$ denotes the lapse function.
In this  simple cosmological set-up the classical dynamics is
defined by the Hamiltonian:
\begin{equation}\label{hham}
 H= N\bigg(-\frac{3}{8\pi G\gamma^2}\;\beta^2\;v + \frac{p_{\phi}^2}{2\,
  v}\bigg),
\end{equation}
where $(\beta, v, \phi,p_{\phi})$ are the kinematical phase space
variables and $\gamma$ is the so-called Barbero-Immirzi parameter.
In terms of scale factor one has $\beta=\gamma\dot{a}/|a|$,
$v=a^3$, and $p_{\phi}$ is the momentum of the field $\phi$. This
model of the universe unavoidably includes the initial
cosmological singularity and has been intensively studied recently
within LQC.

Loop geometry is based on the use of a canonical pair `holonomy
and flux' instead of a canonical pair `connection and triad'
\cite{Ashtekar:2003hd,Bojowald:2006da}. The mapping from
`connection' to `holonomy' is not invertible. The holonomy
function $h_k^{(\lambda)}$ (along straight line of coordinate
length proportional to $\lambda/|a|$) in the fundamental, j = 1/2,
representation of SU(2) group reads \cite{Ashtekar:2006wn}
\begin{equation}\label{hol}
h^{(\lambda)}_k (\beta) = \cos (\lambda \beta/2)\;\id + 2\,\sin
(\lambda \beta/2)\;\tau_k,
\end{equation}
where $\tau_k = -i \sigma_k/2\;$ ($\sigma_k$ are the Pauli spin
matrices). It transforms the gravitational part of the Hamiltonian
of the FRW model into \cite{Ashtekar:2006wn}
\begin{equation}\label{hamR}
    H_g = \lim_{\lambda\rightarrow \,0}\; H^{(\lambda)}_g ,
\end{equation}
where
\begin{eqnarray}\nonumber
H^{(\lambda)}_g = - \frac{v}{2\pi G \gamma^3 \lambda^3}
\sum_{ijk}\,N\, \varepsilon^{ijk}\, Tr \Big(h^{(\lambda)}_i
h^{(\lambda)}_j  \\
\label{hamL} \times (h^{(\lambda)}_i)^{-1}(h^{(\lambda)}_j)^{-1}
h_k^{(\lambda)}\{(h_k^{(\lambda)})^{-1},v\}\Big),
\end{eqnarray}
and where $\varepsilon^{ijk}$ is the alternating tensor. The
Poisson bracket is defined to be
\begin{eqnarray}\nonumber
    \{\cdot,\cdot\}:= 4\pi G\gamma\;\bigg[ \frac{\partial \cdot}
    {\partial \beta} \frac{\partial \cdot}{\partial v} -
     \frac{\partial \cdot}{\partial v} \frac{\partial \cdot}{\partial
     \beta}\bigg]\\\label{PR}
      +
     \frac{\partial \cdot}{\partial \phi} \frac{\partial \cdot}{\partial p_\phi} -
     \frac{\partial \cdot}{\partial p_\phi} \frac{\partial \cdot}{\partial
     \phi}.
\end{eqnarray}

Roughly speaking, quantization of $H^{(\lambda)} :=
H^{(\lambda)}_g + H_\phi$ (where $H_\phi$ is the matter part of
the Hamiltonian) for fixed value of $\lambda$ is the essence of
LQC
\cite{Ashtekar:2006rx,Ashtekar:2006uz,Ashtekar:2006wn,Ashtekar:2007em}.
It means that $H_g$ has been approximated by $H^{(\lambda)}_g$,
since $\lambda$ is assumed to be different from zero. First, one
quantizes the whole phase space, i.e. assigns quantum operators to
the variables $(\beta,v,\phi,p_\phi)$ which act in the kinematical
Hilbert space (KHS) \cite{r1,Dzierzak:2008dy}. Next, one solves
the operator equation
\begin{equation}\label{dirac}
    \widehat{H}^{(\lambda)}\psi=0,
\end{equation}
corresponding to the classical constraint equation $H^{(\lambda)}
= 0$. The task of solving the above equation is far from trivial.
Usually, the space of solutions, $\mathcal{F}$, is not  contained
in KHS. To construct the physical Hilbert space (PHS) one uses the
dual space to $\mathcal{F}$ and special techniques called the
group-averaging method \cite{Marolf:2000iq,Ashtekar:1995zh}. All
that brings about a lot of analytical and numerical work even for
the simple cosmological model considered here.

In what follows we  present the results obtained by using our
quantization method. Making use of (\ref{hol}) in (\ref{hamL})
leads directly to the modified total Hamiltonian corresponding to
(\ref{hham}) given by
 \begin{equation}\label{ham}
 H^{(\lambda)}= N\bigg(-\frac{3}{8\pi G \gamma^2}\;\frac{\sin^2(\lambda
\beta)}{\lambda^2}\;v + \frac{p_{\phi}^2}{2\, v}\bigg)
\end{equation}
The new Hamiltonian is bounded, as a function of $\beta$, and
leads to a modified, singularity-free classical dynamics.
Although, $H^{(\lambda)}$ may generate dynamics in the whole phase
space, the physical sector is constrained to the surface
$H^{(\lambda)} = 0$.

By an observable we mean a function on the phase space which has
vanishing Poisson bracket with the Hamiltonian. We call it an
elementary if it cannot be expressed as a non-invertible function
of another observable. For simplicity of calculations we fix the
gauge by setting
\begin{equation}\label{ggg}
N^{-1}:=\frac{3}{8 \pi G \gamma^2 v} \;\Big(\kappa \gamma |p_\phi|
+ v\,\frac{|\sin(\lambda \beta)|}{\lambda}\Big),
\end{equation}
where $\kappa^2 \equiv 4\pi G/3$. Consequently, (\ref{ham})
reduces to
\begin{equation}\label{hamilton2}
     H^{(\lambda)}= \kappa
\gamma |p_\phi| - v\, \frac{|\sin(\lambda \beta)|}{\lambda} .
\end{equation}
 To identify all observables of our system, we solve the equation
\begin{equation}\label{obsham}
\{\mathcal{O}_j, H^{(\lambda)}\}=0.
\end{equation}
We find that all possible functionally independent elementary
observables are \cite{Dzierzak:2009ip}
\begin{eqnarray}\nonumber
\mathcal{O}_1:= p_{\phi},~~~~~~~~~~~~~~~~~~~~~~~~~\\
\nonumber \mathcal{O}_2:= \phi -
\frac{\textrm{sgn}(p_{\phi})}{3\kappa}\
\textrm{arth}\big(\cos(\lambda \beta)\big),\\
\label{set} \mathcal{O}_3:= \textrm{sgn}(p_{\phi}) \,v\,
\frac{\sin(\lambda \beta)}{\lambda}.~~~~~~
\end{eqnarray}
They satisfy the Lie algebra
\begin{eqnarray}\nonumber
\{\mathcal{O}_2,\mathcal{O}_1\}= 1,~~~\{\mathcal{O}_1,\mathcal{O}_3\}= 0,\\
\label{algebra} \;\{\mathcal{O}_2,\mathcal{O}_3\}= \gamma \kappa
.~~~~~~~~
\end{eqnarray}
For $p_\phi =0$ the algebra (\ref{algebra}) is not well defined,
but we make an extension of it to include this case. The
constraint equation, $H^{(\lambda)} = 0$, takes the simple form
\begin{equation}\label{const}
    \gamma \kappa \,\mathcal{O}_1 =\mathcal{O}_3 .
\end{equation}
Eliminating $\mathcal{O}_3$ from the algebra (\ref{algebra}), by
using (\ref{const}), leads finally to a very simple algebra for
just two variables
\begin{equation}\label{final}
    \{\mathcal{O}_2,\mathcal{O}_1\}= 1,
\end{equation}
where $\mathcal{O}_1, \mathcal{O}_2 \in \mathbb{R}$.

\noindent{\it 3. Energy density and volume operators.} Since
$\dot{\phi}:=\{\phi,H^{(\lambda)}\}= \kappa\gamma\,
\textrm{sgn}(p_{\phi})$ is positive or negative (for $p_{\phi}\neq
0$), $\phi$ changes monotonically so it can be used as an
evolution parameter. To find an evolution of $v$ in terms of
$\phi$, we consider the equation
\begin{equation}\label{v1}
    \frac{dv}{d\phi}=\frac{\dot{v}}{\dot{\phi}}=\frac{\{v,H^{(\lambda)}\}}
    {\{\phi,H^{(\lambda)}\}},
\end{equation}
which in rewritten form is
\begin{equation}\label{v2}
    \frac{\textrm{sgn}(\sin(\lambda \beta))}{\cos(\lambda
\beta)}\;\frac{dv}{v} = 3\kappa\;\textrm{sgn}(p_{\phi})\;d\phi.
\end{equation}
Solution to (\ref{v2}) in terms of elementary observables reads
\cite{Dzierzak:2009ip}
\begin{equation}\label{vol}
    v(\phi) = \kappa\gamma\lambda\,
    |\mathcal{O}_1|\,\cosh\big(3\kappa  (\phi-
    \mathcal{O}_2)\big).
\end{equation}
Taking into account that the energy density of the scalar field is
given by $\rho=p_{\phi}^2/2 v^2$, we get
\begin{equation}\label{rho}
    \rho(\phi)=\frac{1}{2(\kappa\gamma\lambda)^2\cosh^2\big(3\kappa  (\phi-
    \mathcal{O}_2)\big)}.
\end{equation}
The bounce occurs at the maximum of the energy density
\begin{equation}\label{max}
\rho_{\max} = \frac{1}{2}\frac{1}{(\kappa \gamma \lambda)^2}.
\end{equation}

The expressions (\ref{vol}) and (\ref{rho}) show that $v$ and
$\rho$ may be interpreted as a family of  observables labelled by
$\phi$. The physical phase space is now parametrized only by
$\mathcal{O}_1$ and $\mathcal{O}_2$. The $\phi$ variable, an
evolution parameter, does not belong to the physical phase space.
Thus, it will stay classical during the quantization process.
Instead of $\phi$, we may use any evolution parameter specified by
the choice of gauge $N$ in (\ref{ham}). Such a possibility always
exists in the case of globally hyperbolic spacetimes.  In LQC,
contrary to our method, $\phi$ is a phase space variable so it
must be quantized
\cite{Ashtekar:2006rx,Ashtekar:2006uz,Ashtekar:2006wn,Ashtekar:2007em}.
Being a quantum variable it may fluctuate so its use in LQC as an
evolution parameter at the quantum level has poor interpretation.

\noindent{\it 3. Quantum dynamics.} Contrary to the Dirac method,
our quantization method of constrained systems is simple enough to
be fully controlled analytically. In the Schr\"{o}dinger
representation (since $\mathcal{O}_1, \mathcal{O}_2 \in
\mathbb{R}$)  we have
\begin{equation}\label{quant1}
    \mathcal{O}_2 \mapsto \widehat{\mathcal{O}}_2:=x~~,
    ~~\mathcal{O}_1 \mapsto
    \widehat{\mathcal{O}}_1:= -\imath\hbar\partial_x ,
\end{equation}
where $x\in \mathbb{R}$.  The representation of (\ref{final})
defined in the Hilbert space $\mathbb{L}^2(\mathbb{R})$ reads
\begin{equation}\label{quant2}
   [\widehat{\mathcal{O}}_2,\widehat{\mathcal{O}}_1]=\imath\hbar.
\end{equation}
In this representation the energy density operator takes the very
simple form
\begin{equation}\label{quant3}
    \rho\rightarrow \widehat{\rho}:=\frac{1}
    {2(\kappa\gamma\lambda)^2\cosh^2[3\kappa(\phi-x)]}.
\end{equation}
Solution to the eigenvalue problem
\begin{equation}\label{eigen}
   \widehat{\rho}\psi=\rho(x_0)\psi
\end{equation}
for fixed value of $\phi$ reads
\begin{equation}\label{quant4}
    \psi_1=\delta(x-
   x_0)~,~\psi_2=\delta(x+
   x_0-2\phi).
\end{equation}
The eigenvectors (\ref{quant4}) are generalized vectors.  The
spectrum $(0,\frac{1}{2(\kappa\gamma\lambda)^2})$ is doubly
degenerate since $cosh(\cdot)$ is a symmetric function.

Since the evolution (parametrized by $\phi$) of the eigenvalue
corresponding to the generalized eigenvector $\delta(x-x_0)$ reads
\begin{equation}\label{evol}
\widehat{\rho}~\delta(x-x_0)=\frac{\delta(x-x_0)}{2(\kappa\gamma
\lambda)^2\cosh^2\big(3\kappa(\phi-x_0)\big)},
\end{equation}
we conclude that the evolution of the energy density  is the same
as the classical one (\ref{rho}). It is expected that  the
gaussian states, approximating generalized vectors,  have similar
properties.

The resolution of the initial singularity proposed within LQC
\cite{Ashtekar:2003hd,Bojowald:2006da,Ashtekar:2006rx,Ashtekar:2006uz,
Ashtekar:2006wn,Ashtekar:2007em} has been obtained due to the
import of the discreteness of the kinematical geometrical
operators from LQG to LQC \cite{Dzierzak:2008dy}. However, the
kinematical discreteness does not necessarily extend to the
physical Hilbert space, which lies outside of the kinematical
Hilbert space (see comments following (\ref{dirac})). It is
straightforward to show that the claim does not hold in the model
considered here.

The quantum volume operator corresponding to (\ref{vol}) may be
defined as follows \cite{r2}
\begin{eqnarray}\nonumber
    \widehat{v}=\kappa\gamma\lambda\,\frac{1}{2}\,\bigg|
    \widehat{\mathcal{O}}_1\,\cosh[3\kappa  (\phi-
    \widehat{\mathcal{O}}_2)]\\
    \nonumber +\cosh[3\kappa  (\phi-
    \widehat{\mathcal{O}}_2)]\widehat{\mathcal{O}}_1\bigg|\\ \nonumber
    =\kappa\gamma\lambda\hbar\bigg|-\frac{3}{2}\kappa\sinh[3\kappa(\phi-x)]\\
    \label{quant5}+\cosh[3\kappa(\phi-x)]\partial_x\bigg|.
\end{eqnarray}
It is not difficult to find that the eigenvalue problem for the
operator $\widehat{v}$ (for a fixed value of $\phi$) has the
solution
\begin{equation}\label{quant6}
    \widehat{v}\psi=|v|\psi,
\end{equation}
\begin{equation}\label{quant7}
    \psi=\frac{\sqrt{\frac{3\kappa}{\pi}}\exp(i\,\frac{2v}{3\kappa^2
    \gamma\lambda\hbar}\arctan
    e^{3\kappa(\phi-x)})}{\cosh^{\frac{1}{2}}[3\kappa(\phi-x)]},
\end{equation}
where $v\in \mathbb{R}$ and $\psi\in\mathbb{L}^2(\mathbb{R})$ is
normalized.

The spectrum of the volume operator $\widehat{v}$ appears to be
continuous, but our recent analyzes has shown that it is discrete
\cite{Malkiewicz:2009xz}.

\noindent{\it 4. Conclusions.} The global hyperbolicity of
spacetime enables identification of an evolution parameter (e.g.
$\phi$) at the classical level. In our scheme its use has been
extended to the quantum level. Such procedure is possible because
we do not quantize the constraint $H^{(\lambda)} = 0$, but the set
of observables which does not include $\phi$.

The energy scale specific to the Big Bounce can be determined from
(\ref{max}) (quantum and classical energy densities coincide due
to (\ref{quant3}) and (\ref{rho})), but it is not unique because
$\lambda$ is a free parameter of the formalism.

The parameter $\lambda$ has been fixed in LQC by using a discrete
spectrum of the kinematical area operator of LQG. It is an
assumption of LQC  which leads to the commonly expected result
that the Big Bounce transition occurs at the Planck scale
\cite{Ashtekar:2006rx,Ashtekar:2006uz,Ashtekar:2006wn,Ashtekar:2007em}.
It is argued in \cite{Dzierzak:2008dy} and \cite{Bojowald:2008ik}
that this assumption has poor physical justification. In fact,  an
association of the Big Bounce with the Planck scale (within loop
cosmology) may be done easily. If we fix suitably the value of
$\lambda$, our results may fit the Planck scale too: substituting
$\lambda = l_{Pl}$ into (\ref{max}) gives $\rho_{max} \simeq
2,07\;\rho_{Pl}$, and taking $\lambda = 1,44\;l_{Pl}$ leads to
$\rho_{max} \simeq \rho_{Pl}$ ($l_{Pl}$ and $\rho_{Pl}$ denote the
Planck length and energy density, respectively, and we use
$\gamma\simeq 0.24$ determined in the black hole entropy
calculations \cite{Domagala:2004jt,Meissner:2004ju}). However,
real challenge is finding a sound physical justification for the
specific choice of $\lambda$.

The spectrum of primordial gravitational waves is expected to be
sensitive to the holonomy corrections  of the loop cosmology (see,
e.g., \cite{Grain:2009kw,Calcagni:2008ig,Mielczarek:2008pf}).
Detection of the cosmological tensor perturbations  may help to
determine $\lambda$ and identify the energy scale of the Big
Bounce.

Our quantization method, which we applied to FRW model, may be
extended to other cosmologies including simple homogeneous (e.g.
Bianchi I) and isotropic (e.g. Lema\^{i}tre-Tolman) models. Our
next paper will concern the Bianchi I universe.

\noindent {\it Acknowledgments.} We are grateful to Tomasz Bulik,
Piotr  Dzier\.{z}ak and Micha{\l} Ostrowski for helpful
discussions.


\begin{thebibliography}{99}

 \bibitem{PAM} P. A. M. Dirac, \textit{Lectures on Quantum Mechanics}
(New York: Belfer Graduate School of Science Monographs Series,
1964).

\bibitem{NMJ} N. M. J. Woodhouse, \textit{Geometric Quantization}
(New York: Oxford University Press, 1992).

\bibitem{CR} C. Rovelli \textit{Quantum Gravity}
(Cambridge: Cambridge University Press, 2004).

\bibitem{TT} T. Thiemann \textit{Modern Canonical Quantum General Relativity}
(Cambridge: Cambridge University Press, 2007).

\bibitem{Ashtekar:2003hd}
  A.~Ashtekar, M.~Bojowald and J.~Lewandowski,
  ``Mathematical structure of loop quantum cosmology'',
  Adv.\ Theor.\ Math.\ Phys.\  {\bf 7}, 233 (2003) .
  [arXiv:gr-qc/0304074].

\bibitem{Bojowald:2006da}
  M.~Bojowald,
  ``Loop quantum cosmology'',
  Living Rev.\ Rel.\  {\bf 8}, 11 (2005)
  [arXiv:gr-qc/0601085].

\bibitem{Ashtekar:2006rx}
  A.~Ashtekar, T.~Pawlowski and P.~Singh,
  ``Quantum nature of the big bang'',
  Phys.\ Rev.\ Lett.\  {\bf 96}, 141301 (2006).
  [arXiv:gr-qc/0602086].

\bibitem{Ashtekar:2006uz}
  A.~Ashtekar, T.~Paw{\l}owski and P.~Singh,
  ``Quantum nature of the big bang: An analytical and numerical
  investigation'',
  Phys.\ Rev.\  D {\bf 73} 124038 (2006).
  [arXiv:gr-qc/0604013].

\bibitem{Ashtekar:2006wn}
  A.~Ashtekar, T.~Paw{\l}owski and P.~Singh,
  ``Quantum nature of the big bang: Improved dynamics'',
  Phys.\ Rev.\  D {\bf 74}, 084003 (2006).
  [arXiv:gr-qc/0607039].

\bibitem{Ashtekar:2007em}
  A.~Ashtekar, A.~Corichi and P.~Singh,
  ``On the robustness of key features of loop quantum cosmology'',
  Phys.\ Rev.\  D {\bf 77}, 024046 (2008).
  [arXiv:0710.3565 [gr-qc]].


\bibitem{r1} Strictly speaking, in LQC one quantizes $\exp{(i \lambda
\beta)}$ instead of $\beta$.

\bibitem{Dzierzak:2008dy}
  P.~Dzierzak, J.~Jezierski, P.~Malkiewicz and W.~Piechocki,
  ``Conceptual issues concerning the Big Bounce'',
  arXiv:0810.3172.


\bibitem{Marolf:2000iq}
  D.~Marolf,
  ``Group averaging and refined algebraic quantization: Where are we
  now?'',
  arXiv:gr-qc/0011112.

\bibitem{Ashtekar:1995zh}
  A.~Ashtekar, J.~Lewandowski, D.~Marolf, J.~Mourao and T.~Thiemann,
  ``Quantization of diffeomorphism invariant theories of connections with local
  degrees of freedom'',
  J.\ Math.\ Phys.\  {\bf 36}, 6456 (1995).
  [arXiv:gr-qc/9504018].

\bibitem{Dzierzak:2009ip}
  P.~Dzierzak, P.~Malkiewicz and W.~Piechocki,
  ``Turning Big Bang into Big Bounce: I. Classical Dynamics,''
  arXiv:0907.3436 [gr-qc].


\bibitem{r2} We define the modulus of a self-adjoint operator $\mathcal{O}$ as
follows: If $\mathcal{O} f_a = a f_a$, then $|\mathcal{O}| f_a :=
|a| f_a$.

\bibitem{Malkiewicz:2009xz}
  P.~Malkiewicz and W.~Piechocki,
  ``Foamy structure of spacetime,''
  arXiv:0907.4647 [gr-qc].


\bibitem{Bojowald:2008ik}
  M.~Bojowald,
  ``Consistent Loop Quantum Cosmology'',
  Class.\ Quant.\ Grav.\  {\bf 26} 075020 (2009).

\bibitem{Domagala:2004jt}
  M.~Domagala and J.~Lewandowski,
  ``Black hole entropy from quantum geometry'',
  Class.\ Quant.\ Grav.\  {\bf 21} 5233 (2004).

\bibitem{Meissner:2004ju}
  K.~A.~Meissner,
  ``Black hole entropy in loop quantum gravity'',
  Class.\ Quant.\ Grav.\  {\bf 21} 5245 (2004).


\bibitem{Grain:2009kw}
  J.~Grain and A.~Barrau,
  ``Cosmological footprints of loop quantum gravity'',
  arXiv:0902.0145.

\bibitem{Calcagni:2008ig}
  G.~Calcagni and G.~M.~Hossain,
  ``Loop quantum cosmology and tensor perturbations in the early
  universe'',
  arXiv:0810.4330.

\bibitem{Mielczarek:2008pf}
  J.~Mielczarek,
  ``Gravitational waves from the Big Bounce'',
  JCAP {\bf 0811}, 011 (2008).
  [arXiv:0807.0712 [gr-qc]].



\end{thebibliography}
\end{document}